# Assortative pairing alone can lead to a structured biota in organisms with cultural transmission

Petr Tureček[1,2,*], Michal Kozák[3], Jakub Slavík[4]

[1]Department of Philosophy and History of Science, Faculty of Science, Charles University; Prague 2, 128 00, Czech Republic.

[2]Center for Theoretical Study, Charles University and Czech Academy of Sciences, Jilská 1, Prague 1, 110 00, Czech Republic.

[3]Department of Mathematics, Faculty of Nuclear Sciences and Physical Engineering, Czech Technical University in Prague, Trojanova 13, 120 00, Prague 2, Czech Republic

[4]Institute of Information Theory and Automation, The Czech Academy of Sciences, Pod Vodárenskou věží 4, 180 00, Prague 8, Czech Republic

*petr.turecek@natur.cuni.cz

**Abstract**

Spatial separation is often included in models of ethnic divergence but it has also been realised that urban subcultures can, and frequently do, emerge in sympatry. Previous research tended to attribute this phenomenon to the human tendency to imitate self-similar individuals and actively differentiate oneself from individuals recognized as members of an outgroup. Application of such a model to non-human animals has been, however, viewed as problematic. We present a parsimonious model of subculture emergence where the algorithm of social learning does not require the assumption of an 'imitation threshold'. All it takes is a slight modification of Galton-Pearson's biometric model previously used to approximate cultural inheritance. The new model includes proportionality between the variance of inputs (cultural 'parents') and the variance of outputs (cultural 'offspring'). In this model, assortment alone can lead to the formation of distinct cohesive clusters of individuals (subcultures) with a low within-group and large between-group variability even in absence of a spatial separation or disruptive natural selection. Sympatric emergence of arbitrary behavioural varieties preceding ecological divergence may thus represent the norm, not the exception, in all cultural animals.

## 1 Introduction

There is a long-standing debate about the possibility of speciation that is sympatric, i.e., the emergence of multiple species from one species without geographic isolation[1–8]. Although the importance of the geographical aspect of speciation had lately been questioned[9,10], spatial isolation is still believed to play a pivotal role in divergent evolution because it limits the gene flow between adjacent populations[11]. In this contribution, we investigate the implications of assortative mating within a model of nonparticulate multidimensional inheritance. The inheritance model can be used to approximate both an additive genetic transmission (Galton-Pearson model[12]) or a cultural one (Parental Variability-Dependent Inheritance[13]). We demonstrate that in a model where the effect of culture is dominant, assortment can even on its own, i.e., in absence of any disruptive natural selection, lead to the formation of distinct cohesive clusters of individuals with low within-group and large between-group variance. Parental Variability-Dependent Inheritance can explain the frequent sympatric emergence of subcultures and varieties characterised by limited gene flow (although the subcultures or varieties can diverge ecologically at a later point). This model might be particularly relevant for speciation in taxa with a high capacity for cultural evolution, such as the cetaceans[14], songbirds[15,16], or hominins[17], which are known for their potential to undergo adaptive ecological radiation.

**Darwin's mystery of mysteries**

Originally, the famous designation 'mystery of mysteries'[18] was not reserved just for the formation of distinct species. It referred to the emergence of distinct groups of organisms with high between-group and low within-group variation[19]. Darwin stressed the absence of fundamental differences between varieties, subspecies, species, and higher taxa[20]. Differences between such categories were supposed to be the question of scale rather than quality. Current research tends to support this view[9,21]. It seems that reproductive isolation can emerge quite easily even in conditions characterised by an uninterrupted gene-flow[7]. Two distinct varieties may develop into two separate species via several broadly overlapping stages: (1) homogeneous gene flow between populations, (2) heterogeneous gene flow where alleles directly linked to diverging traits rarely cross the boundaries between populations, and finally (3) separate species that meet the condition of full reproductive isolation[22]. It should be noted, however, that resistance from the mainstream against this view is still quite considerable[11].

Nevertheless, the subject of structured biota is no less challenging than the better-known species-oriented perspective[19]. Why do, for example, distinct cultures, subcultures, or ethnic groups exist? This is difficult to answer because research in cultural evolution tends to focus on arbitrary signs or ornaments which do not necessarily have an adaptive function that would allow for an ecological explanation of the divergence[23]. Computer simulations of minimal counterfactual systems can shed light on this conundrum.

**Nonparticulate inheritance models**

Two models of nonparticulate inheritance have recently been employed in studies of cultural evolution[12,13]. These models do not rely on, in this context restrictive, beliefs about additional genetic variation: they can be viewed as general inheritance models such as those which preceded or existed in parallel with Mendelism. They have been overlooked due to evolutionary biologists' focus on genetic models which are believed to provide a good approximation of the simplest form of organismal inheritance. Both models are based on simple blending inheritance but replace the problematic assumption of the offspring ($t_o$) being exactly in the arithmetic mean ( $\mu(x_1, \ldots, x_M) = \frac{1}{M}\sum_{i=1}^{M} x_i$ ) of parental values ($t_{p1}, t_{p2}$) with the assumption of a random normal distribution of offspring (N($m, V$), with mean $m$ and variance $V$). Galton-Pearson (GP) model (Equation (1.1)) assumes a mutation term characterised by a constant standard deviation ($\eta$) independent of the difference between parental values.

$$t_o = \mu(t_{p1}, t_{p2}) + \text{N}(0, \eta^2), \tag{1.1}$$

This model is based on Galton's experiments with inheritance of the seed size in sweat pea and investigations of heredity of human height. It can very well approximate genetic inheritance if we assume a large population and a high number of freely recombining genes with additive genetic variance[24]. Despite its genetic roots, it has been used in approximations of cultural inheritance of continuous traits[12].

Parental Variability-Dependent Inheritance (PVDI, Equation (1.2)), suggested as an alternative, supposes that the standard deviation ($\sigma(x_1, \ldots, x_M) = \sqrt{\frac{1}{M}\sum_{i=1}^{M}(x_i - \mu(x_1, \ldots, x_M))^2}$ ) of offspring is proportional (with ratio $v$) to the standard deviation of parental trait values, which is in biparental inheritance equal to one-half of the parental distance[13].

$$t_o = \mu(t_{p1}, t_{p2}) + \text{N}\big(0, v^2 \sigma^2(t_{p1}, t_{p2})\big) \tag{1.2}$$

In the PVDI system, homogeneous parents produce homogeneous offspring and heterogeneous parents produce heterogeneous offspring.

Both nonparticulate models from Equations (1.1) and (1.2) can be easily combined and generalised to a multidimensional form (Equation (1.3)), where $A_i = (t_{i,1}, t_{i,2}, \ldots, t_{i,D}) \in \mathbb{R}^D$ represents individual's position in a Euclidean $D$-dimensional trait-space.

$$A_o = \mu(A_{p1}, A_{p2}) + \frac{A_{p1} - A_{p2}}{\|A_{p1} - A_{p2}\|} \text{N}\left(0, \eta^2 + v^2 \frac{\|A_{p1} - A_{p2}\|^2}{4}\right) \tag{1.3}$$

There, the offspring assumes a position on a vector connecting her parents. Her distance from the point in-between parental positions is normally distributed. The lower the proportional offspring variation ($v$) relative to the constant offspring variation ($\eta$), the closer is the model to a pure Galton-Pearson inheritance. A visual comparison of the models can be found in Figure 1.

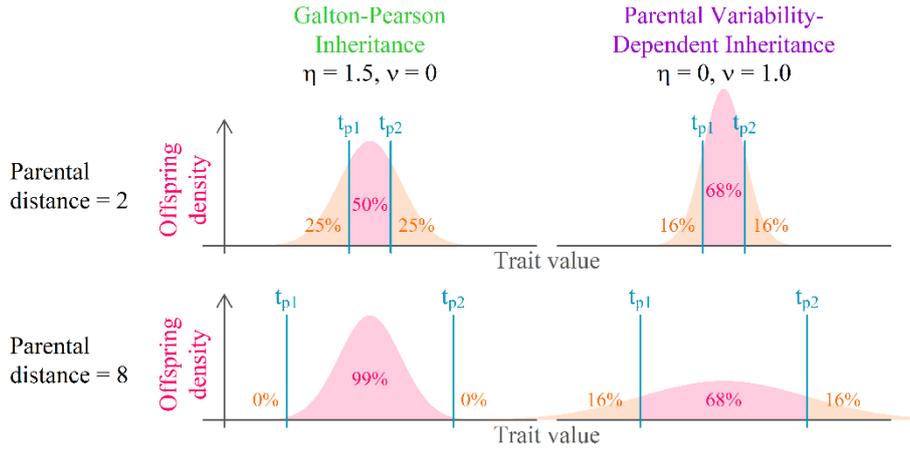

**Figure 1 | Model comparison.** Offspring distribution function is given by the arithmetic mean of parental values and phenotypic mutation (see Equations (1.1) and (1.2)). In a system with pure PVDI, the proportion of offspring between parental values is constant. In a system with GP inheritance, the proportion depends on distance between parental values.

Assortative social learning and mating seems almost omnipresent in human populations[25] and has been abundantly documented also in nonhuman animals[16,26]. It has been previously identified as a necessary but not sufficient condition of sympatric speciation[3].

## 2 Methods
### Overview

We employ an agent-based simulation to study the emergence of a structured biota under the assumption of assortative pairing. Individuals close to each other in a multidimensional trait space are considered similar. The difference between two individuals can be calculated as the distance between their positions in a trait space[27]. We assume no external natural selection acting on the population and there are only relative preferences for pairing between individuals. In our simulations, relative preferences between agents are determined by a homophily coefficient ($h$) which expresses a preference for self-similarity. If $h$ is negative, we observe a negative assortment where dissimilar individuals are more likely to pair. When $h$ is positive, there is a positive assortment and similar individuals pair with a higher probability. For $h = 0$, there is no assortment, and the pairing is completely random.

We can picture this as a cultural transmission or procreation in a multidimensional trait space. In each step, each agent selects an interaction partner ('role model') and modifies her position along the vector between her and her interaction partner. An agent will probably move towards the partner, sometimes overshooting her original position. With the same probability as overshooting her position, the agent moves away from the partner (in Figure 1, $t_{p1}$ can be read as the agent's original position, $t_{p2}$ as the selected role model's position, and offspring density as the agent's resulting position distribution after the interaction).

**The formal model**

In each simulation run, each agent $A_i$ in a population $P$ of size $n$ is represented by its position in a Euclidian $D$-dimensional space (a 'trait space'):

$$P = (A_1, A_2, \ldots, A_n), \; A_i = (t_{i,1}, t_{i,2}, \ldots, t_{i,D}) \in \mathbb{R}^D. \tag{2.4}$$

The difference between any two individuals in a trait space can be calculated as the distance between two points in $D$-dimensional space:

$$|A_i - A_j| = \sqrt{\sum_{d=1}^{D}(t_{i,d} - t_{j,d})^2}. \tag{2.5}$$

In this trait space, we are simulating a process of non-particulate position inheritance. There is no external natural selection operating on the population and only relative preferences for pairing between individuals. In each time step or 'generation' $g$, each agent selects a partner probabilistically. Relative preference of the focal individual $A_i$ for agent $A_j$ is given by

$$\text{rel. pref}(A_i, A_j) = \left(\frac{1}{1 + \|A_i - A_j\|}\right)^h \tag{2.6}$$

where $h$ is a homophily coefficient, that is, a measure of preference for self-similarity. This well-behaved function ensures that self-preference, i.e., preference for individuals whose distance from an agent is zero, serves as a referential value of 1, and other preferences follow proportionally depending on $h$.

The probability that $A_i$ selects $A_j$ is given by

$$\text{prob}(A_i, A_j) = \frac{\text{rel. pref}(A_i, A_j)}{\sum_{k=1}^{n} \text{rel. pref}(A_i, A_k)}. \tag{2.7}$$

If $h$ is negative, we observe heterophily, that is, negative assortative mating where dissimilar individuals pair with a higher probability. For positive $h$, pairing follows homophily, a positive assortative mating, and it is similar individuals that pair with a higher probability. For $h = 0$, there is no assortment, relative preference for all individuals is 1, and pairing is completely random. The absolute value of homophily $|h|$ corresponds to the strength of assortment. For $h = 1$, agent $A_j$ who is twice closer to the focal agent $A_i$ than agent $A_k$ will be selected as a partner by $A_i$ approximately twice as often as $A_k$ (assuming $\|A_i - A_j\|$ and $\|A_i - A_k\|$ are significantly larger than 1). For $h = 2$, $A_j$ will be selected approximately four times as often as $A_k$. For $h = 3$, $A_j$ will be selected approximately eight times as often as $A_k$, etc. Similarly, for $h = -1$, $A_k$ will be selected twice as often as $A_j$ etc. *mutatis mutandis*.

Because each individual $A_i(g)$ at time step $g$ selects one role model $A_j(g)$, we obtain $n$ pairs of parents in all time steps. Each pair of parents creates one offspring, denoted for this purpose $A_i(g + 1)$, whose position in the trait space is given by probabilistic non-particulate inheritance combining Galton-Pearson and PVDI terms.

$$A_i(g+1) = \mu(A_j(g), A_i(g)) + \frac{A_j(g) - A_i(g)}{\|A_j(g) - A_i(g)\|} N\left(0, \eta^2 + \nu^2 \frac{\|A_j(g) - A_i(g)\|^2}{4}\right) \quad (2.8)$$

Agent's position in the following $g + 1$ timestep, $A_i(g + 1)$, therefore always lies on a line connecting the parental points (a variant of the model with additional normal noise can be found in Supplement S4). The distance of $A_i(g + 1)$ from the arithmetic average of $A_i(g)$ and $A_j(g)$ is normally distributed.

The construction of agent configuration in $g + 1$, $P(g + 1)$ from $P(g)$ takes only a single time step, which means that all agents in our simulation alter their positions in synchrony.

A similar model that pairs individuals for each time step anew can also be constructed. With this modification, $n/2$ pairs of agents are formed in a generation $g$, each creating two new agents, ensuring that the population size remains constant, whereby the subsequent time step $g + 1$ follows the same inheritance algorithms. This modification invites new interpretations: (a) vertical cultural transfer from biological parents to biological offspring, meaning that one step is identified with a biological generation, or (b) exclusive interactions, such as discussions, conversations, exchanges of opinions, after which the positions of both interacting individuals change simultaneously. In this model, an additional specification of the algorithm must provide for within-pair exclusivity. The selection of interaction partners in $g$ is decided one agent at a time in a random order. If an agent is already selected as an interaction partner, she does not undertake a partner selection of her own. In any case, both models – let us refer to them as 'inspirational' and 'interactional' respectively – are highly similar and lead to equivalent conclusions. It therefore should not be controversial to derive implications for a vertical cultural transmission from a computer simulation of an inspirational model and *vice versa*. The 'inspirational' model, which is the focus of this manuscript, does not require any additional specifications.

**Computer simulation**

The initial state of the computer simulation was set to a single cloud of normally distributed points with standard deviation $\sigma = 100$ along each trait value. Simulation results were obtained for $D = 10$. Simulation parameters ($n = 250, 500, 1000$, $\eta = 1, 10, 100$, $0 \leq h \leq 4$, $0 \leq \nu \leq 3.0$) were chosen so as to demonstrate important transitions between systems with different frequencies of divergence. From the initial configuration, the partner selection + position inheritance algorithm was iterated for 200 steps in each simulation run (400–800 runs per parameter combination depending on the detail required given the variation in the number of distinct clusters).

**Analysis and visualisation**

The average distance between agents in the trait space was calculated to assess whether the population of agents would collapse into a single point (an equivalent of loss of variability in simulations of one-dimensional cultural adaptation) or enter a feedback loop of permanent expansion and variability boost (an equivalent of variability explosion in simulations of one-dimensional cultural adaptation) in a given simulation run. Effective dimensionality of the trait space was quantified by the proportion of variance of agent positions explained by the first three principal components (PCs) of the $D$-dimensional trait space and by the number of PCs necessary for capturing 99% of all variance. These measures were highly correlated ($cor = 0.9$), so in the following, we only worked with the first measure. The number of distinct clusters was evaluated using the HDBSCAN algorithm (for a description of the algorithm, see https://hdbscan.readthedocs.io/en/latest/how_hdbscan_works.html).

The development of agent network in the trait space is visualised across time steps as a series of static images and as an animation capturing the dynamical process of clustering (see Supplementary animations). The 3D scatterplots work with the first three Principal Components (standardised to have mean=0 and sd=1 along each PC) and summarise the layout of points in $D$-dimensional trait-space. To minimise scatterplot rotation between adjacent frames, PC1, PC2, and PC3 are mapped to axes x, y, and z through PC rearrangement and reversion, such that $\text{cor}(x_g, x_{g+1}) + \text{cor}(y_g, y_{g+1}) + \text{cor}(z_g, z_{g+1})$ is maximised. The points indicating agents' relative positions in PC1–PC3 space are coloured according to their assignment to distinct clusters. The biggest subgroup from cluster Γ in time step $g$ that belongs to a single cluster in time step $g + 1$ inherits the colour of Γ from the previous step. The measure of dimensionality reduction (Figure S1), the number of distinct clusters (Figure 2), and the average distance between agents (Figure S2) are visualised as variables dependent on time in a graphical summary of a single simulation run.

The average values of measures for all simulation runs with the same parameter configuration are visualised as colour shades depending on parameters $n$, $h$, $\nu$, and $\eta$.

A straightforward extension of the model allows us to focus on relative rather than absolute differences between individuals in processes that drive cultural acquisition. In such model, we normalise the positions of agents so as to maintain the average distance between agents constant after each step. Thanks to this additional step, the model leads neither to variability loss (where $\nu$ and $\eta$ are too low) nor to variability explosion (where $\nu$ is too high), which were both present in the original model. In effect, normalisation binds $\eta$ to overall population variance while keeping it independent of the two particular values selected as parental traits. This resembles Fisher's genetic model which reconciled biometric inheritance with Mendelism under the assumption of additive genetic variance[24]. In the main manuscript, we abstained from any extensions, including normalisation, because our aim was to demonstrate the potential of a very simple model relying only on bilateral relationships between interacting agents to sustain a reasonable cultural variation. A more detailed exposition of the normalised model can be found in Supplement S3.

Demonstrative examples were run using R[28]. The code for parallel runs was written in Python[29] using *Numpy*[30] infrastructure and delegated to processor cores using R package *parallel*[28] and *reticulate*[31]. Visualisations were created using the base R graphics (all code is available at: https://osf.io/pvyhe/?view_only=a79f0d07847f45cab069a8b8f09f258b)

## 3 Results

We found that structured biota, manifested as distinct groups with high between-group and low within-group variance, emerges from positive assortment alone when PVDI prevails over GP, that is, when the constant standard deviation $\eta$ in an inheritance model that combines the GP with PVDI is small (Figure 3B). The emergence of distinct subcultures was observed even when – in contrast to some previous studies [32] – we did not assume that agents coordinate their positions below or extremise their positions above certain threshold distance.

In a system with a considerable influence of constant offspring variation, assortment alone cannot lead to the emergence of a structured biota (Figure 2). Instead, the agents form a single uniform cluster (Figure 3A). This outcome is altered neither by homophily nor by population size.

Regardless of the constant offspring standard deviation $\eta$, smaller populations face a reduction of effective trait-space dimensionality (Supplement S1, Figure S1). The effective dimensionality decreases when relative offspring variability $\nu$ is high. In smaller populations, unidimensional polarisation or variability loss is more likely, which conforms to the results of previous research[33]. We thus arrive at an interesting interplay between agent positions and the effective trait space, where one depends on the other and influences it at the same time. Our simulations thus show that the formation of diverse cultural systems, where individuals differ along multiple dimensions, requires large populations.

The inability of the Galton-Pearson system, which approximates the model of polygenic additive inheritance, to form distinct varieties in the presence of assortative pairing alone might be interpreted as providing support for ecological theories of sympatric divergence[7,9,11,18]. In absence of any cultural inheritance, natural and sexual selection must take place concurrently to generate separate ecotypes or species[4,8,18].

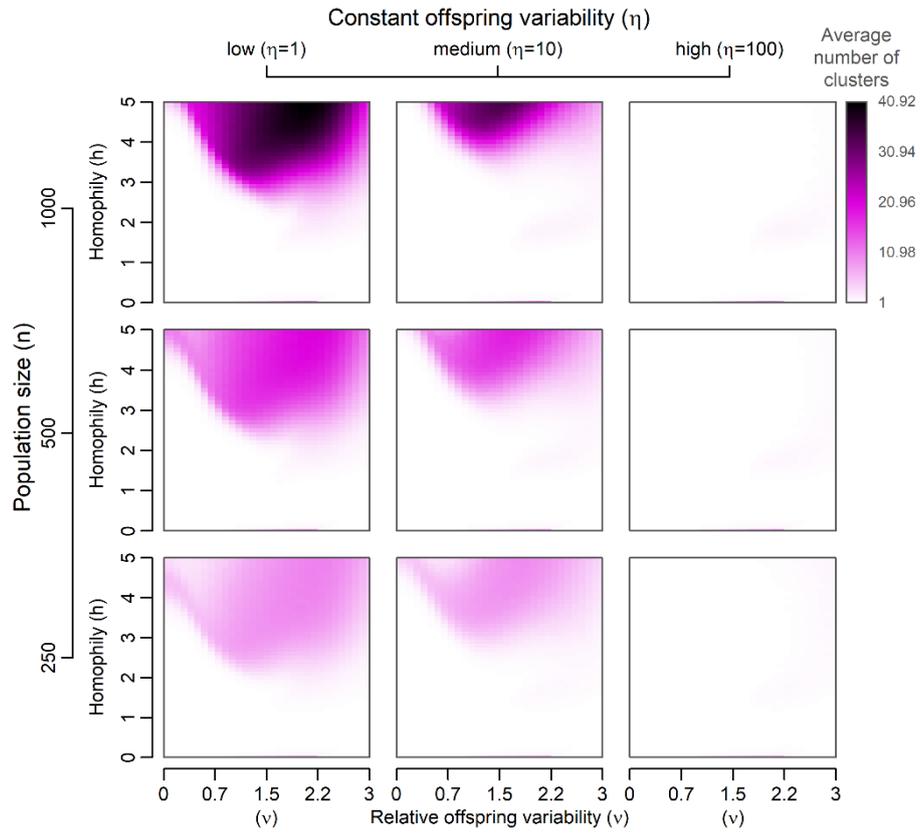

**Figure 2 | A graphical summary of the tendency to form subcultures** after 200 model generations. The points in the 10-dimensional culture space were normally distributed across all dimensions at the beginning of each simulation run, and 800 simulation runs were executed for each parameter combination for $\eta = 1$, 600 runs for $\eta = 10$, and 400 runs for $\eta = 100$.

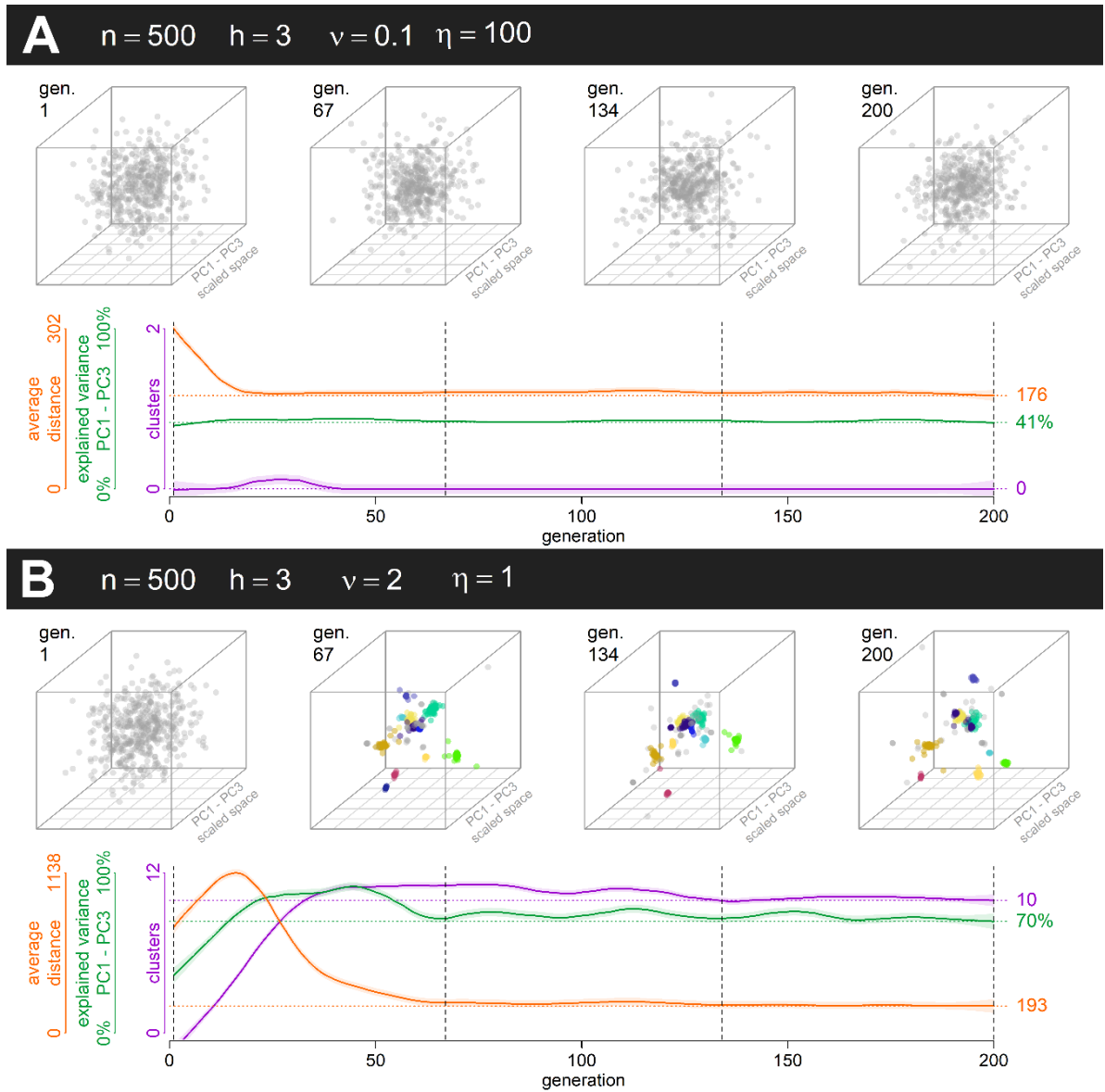

**Figure 3 | Two simulation runs, one of a system strongly influenced by Galton-Pearson inheritance (A) and one strongly influenced by PVDI (B).**
The configuration of points across the first three principal components is displayed at the beginning and after each third of the simulation run. The system with PVDI inhabits the culture space in a discontinuous manner and forms semi-isolated clusters which are stable over time. (See also Supplementary animations 3A and 3B, all supplementary animations are deposited in a separate folder on https://osf.io/pvyhe/?view_only=a79f0d07847f45cab069a8b8f09f258b). Standardised first three Principal Components (PC1–PC3 scaled space; for further elucidation see Methods) rotated to minimise changes between adjacent images are used to visualise the ten-dimensional configuration in 3D scatterplots. (More examples of simulation runs are provided in Supplement S2.)

## 4 Discussion

We argue that unfragmented species of sexually reproducing organisms with inheritance system that follows the PVDI (such as culture) are held together by the same force that holds together the morphological lineages of asexual organisms, that is, by stabilising natural selection. In such species, sympatric speciation should not be viewed as an exception but rather as a norm. Distinct variants that emerge spontaneously can easily settle into distinct ecological niches if there is a potential for divergent selection[14]. Such clusters are precursors of cultures, subcultures, guilds, alternative subsistence strategies, political factions, etc.[34]. We suggest that this process enables organisms with the capacity for social transmission to form and maintain distinct ecotypes defined by rituals, ethnic markers, and dialects even prior to the emergence of any genetic differences between these groups[35]. Such organisms should be therefore capable of exploiting natural resources in ways which are out of the reach of organisms that rely solely on genetic inheritance. Sympatric fission, that is, cases where cultural divergence precedes any spatial dislocation of individuals, may play a vital role in the distribution of cultural variants across time and space[14,34]. Gradually, one may see an accumulation of genetic differences aiding further differentiation between groups[22]. Cultural assortment sparked by preference for self-similar individuals may be the key factor that limits gene flow between emerging subpopulations. The attested historical adaptive radiations of hominins[17] may be in part due to the vast cultural capacity of our lineage.

# Supplementary material

## Contents

All supplementary animations are deposited in a separate folder on https://osf.io/pvyhe/?view_only=a79f0d07847f45cab069a8b8f09f258b.

## Supplement S1. Expected measurement values across the parameter space

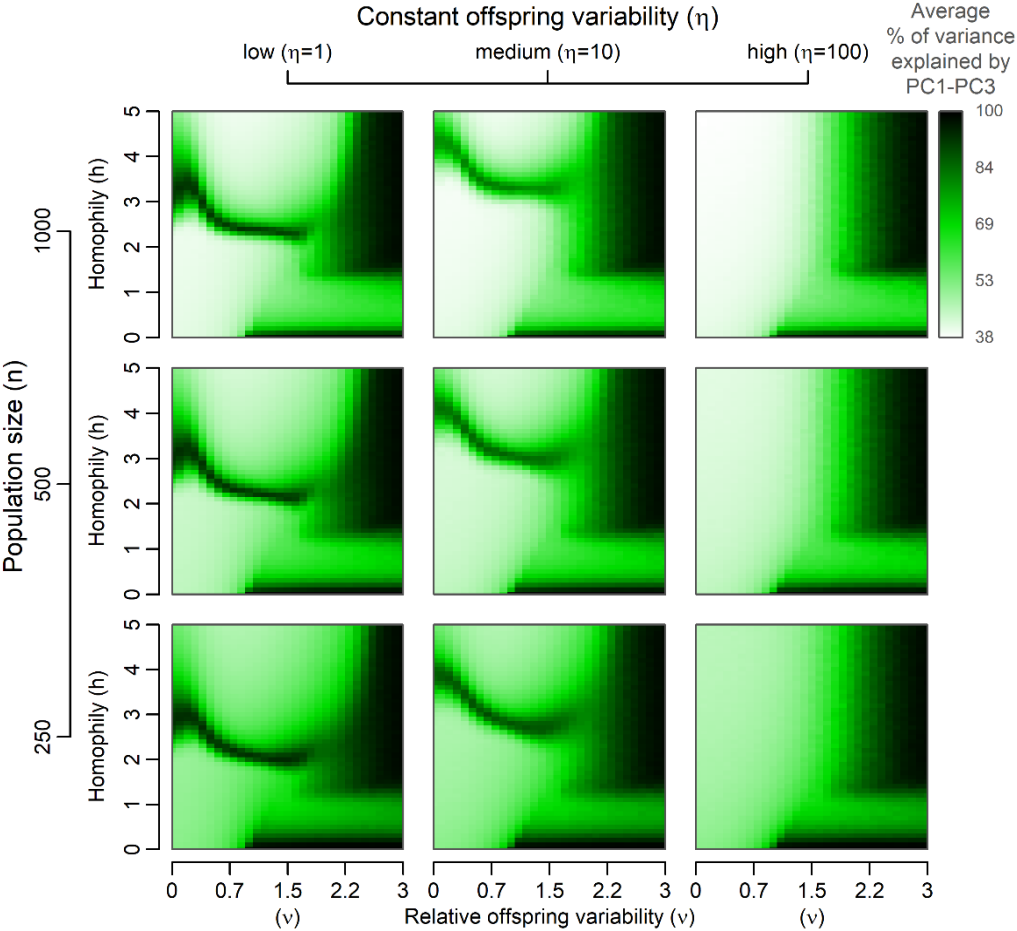

**Figure S1. A graphical summary of expected reduction in the number of effective dimensions** after 200 model generations. The same set of simulation runs as in Figure 2 was used to generate the image. PC1–PC3 stand for the first three Principal Components.

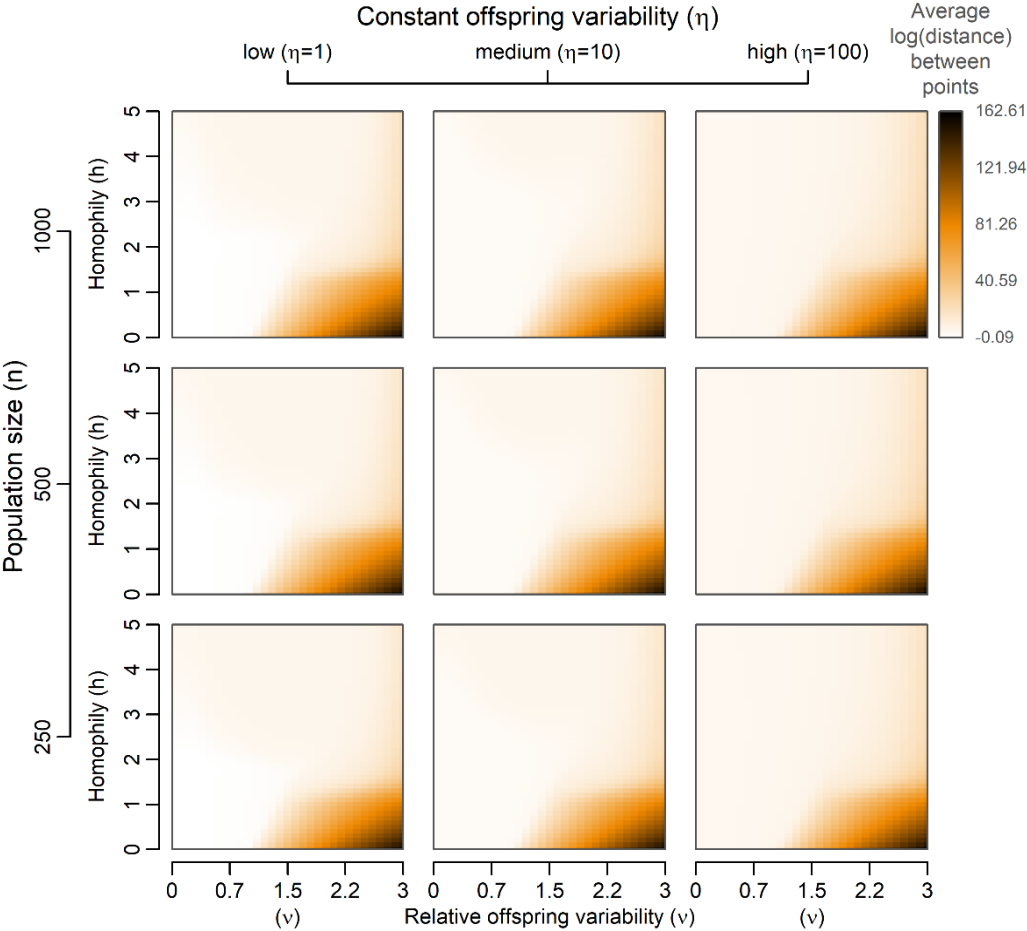

**Figure S2. A graphical summary of the expected average distance between agent positions in a culture-space** after 200 model generations. The same set of simulation runs as in Figure 2 was used to generate the image.

# Supplement S2. Example simulation runs

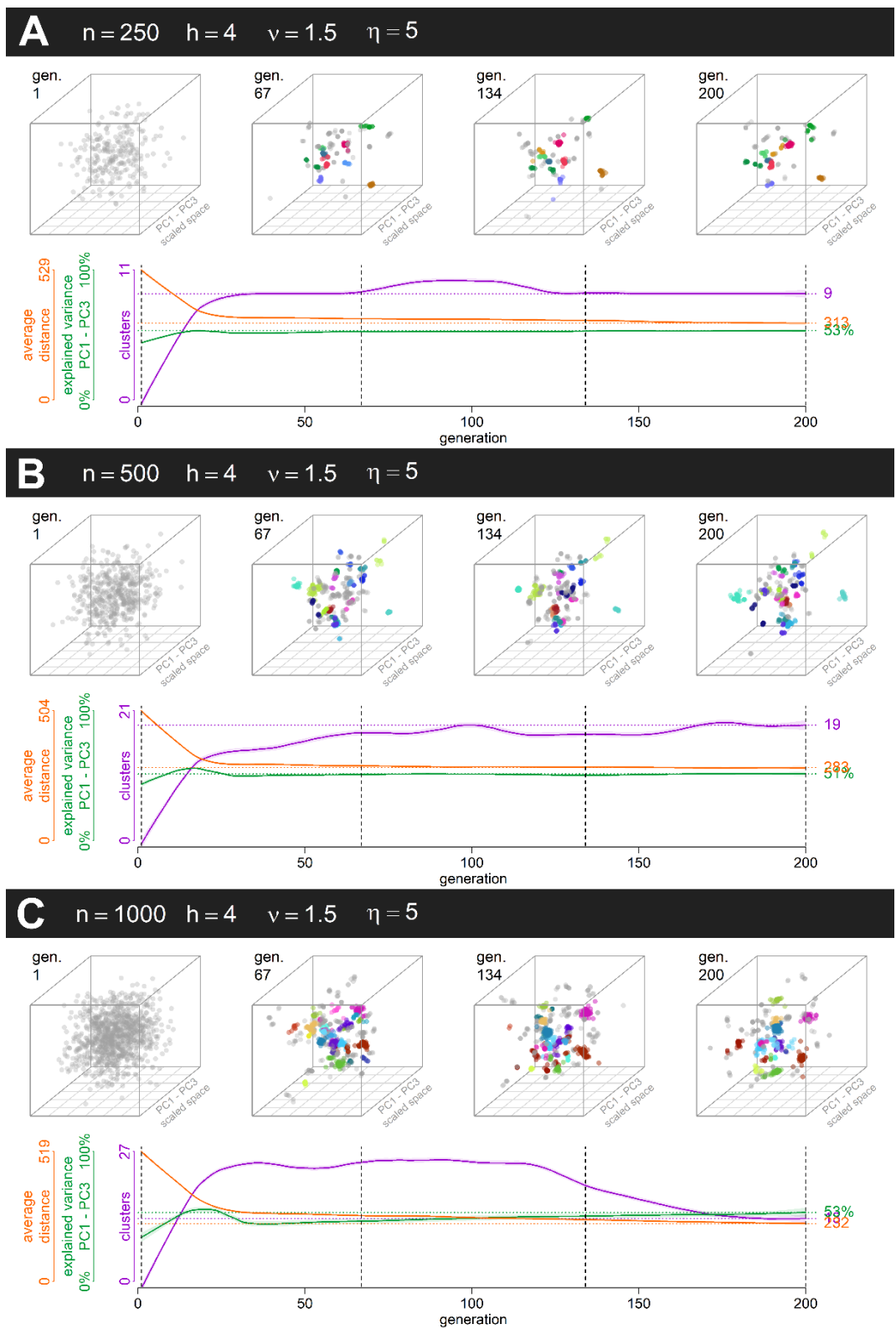

**Figure S3. Example simulations leading to stable distinct clusters, one simulation per small (A: n = 250), medium (B: n = 500), and large (C: n=1000) population.** All other parameters remain constant across the simulation runs (See supplementary animations S3A–S3C).

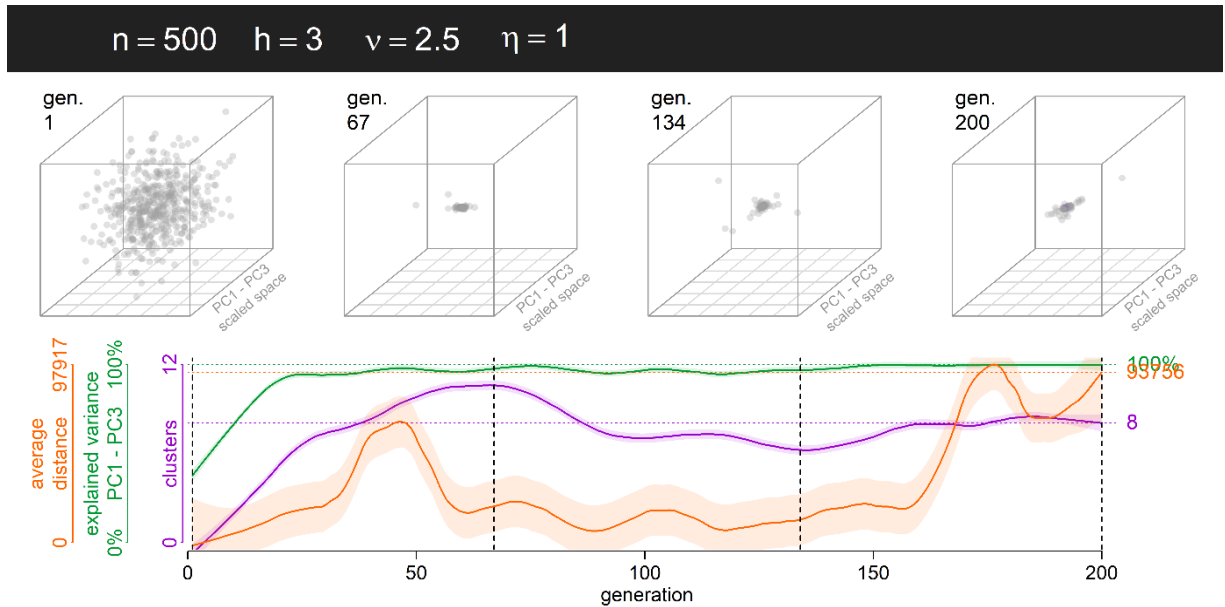

**Figure S4. Example simulation leading to a lower number of distinct clusters due to a lower homophily (h = 3) and larger relative offspring variance (v = 2.4).** (See supplementary animation S4).

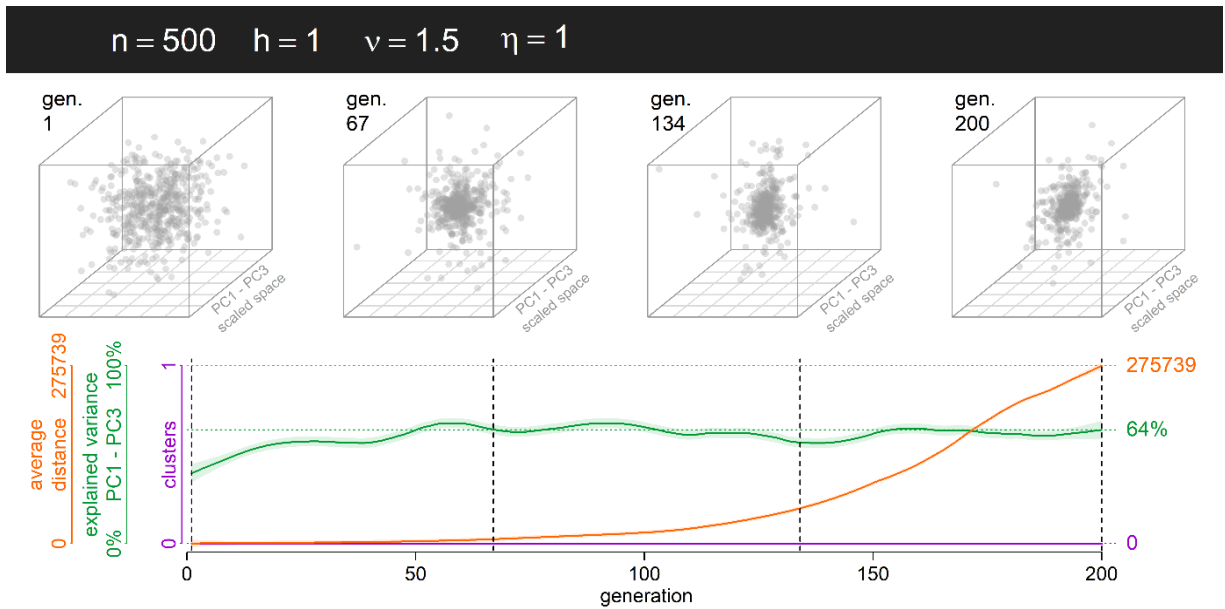

**Figure S5. Example simulation leading to a very low number of distinct clusters due to a weak assortment between agents (h = 1)** (See supplementary animation S5).

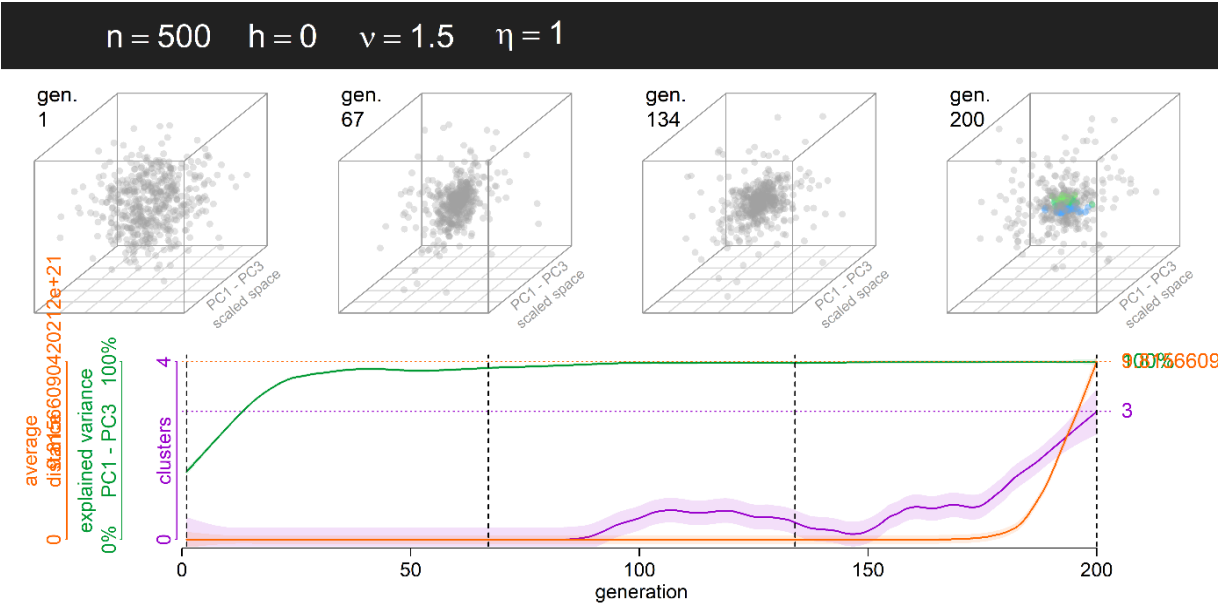

**Figure S6. Example simulation leading to variability explosion and weak clustering along a single dimension due to random interactions between agents (h = 0)** (See supplementary animation S6).

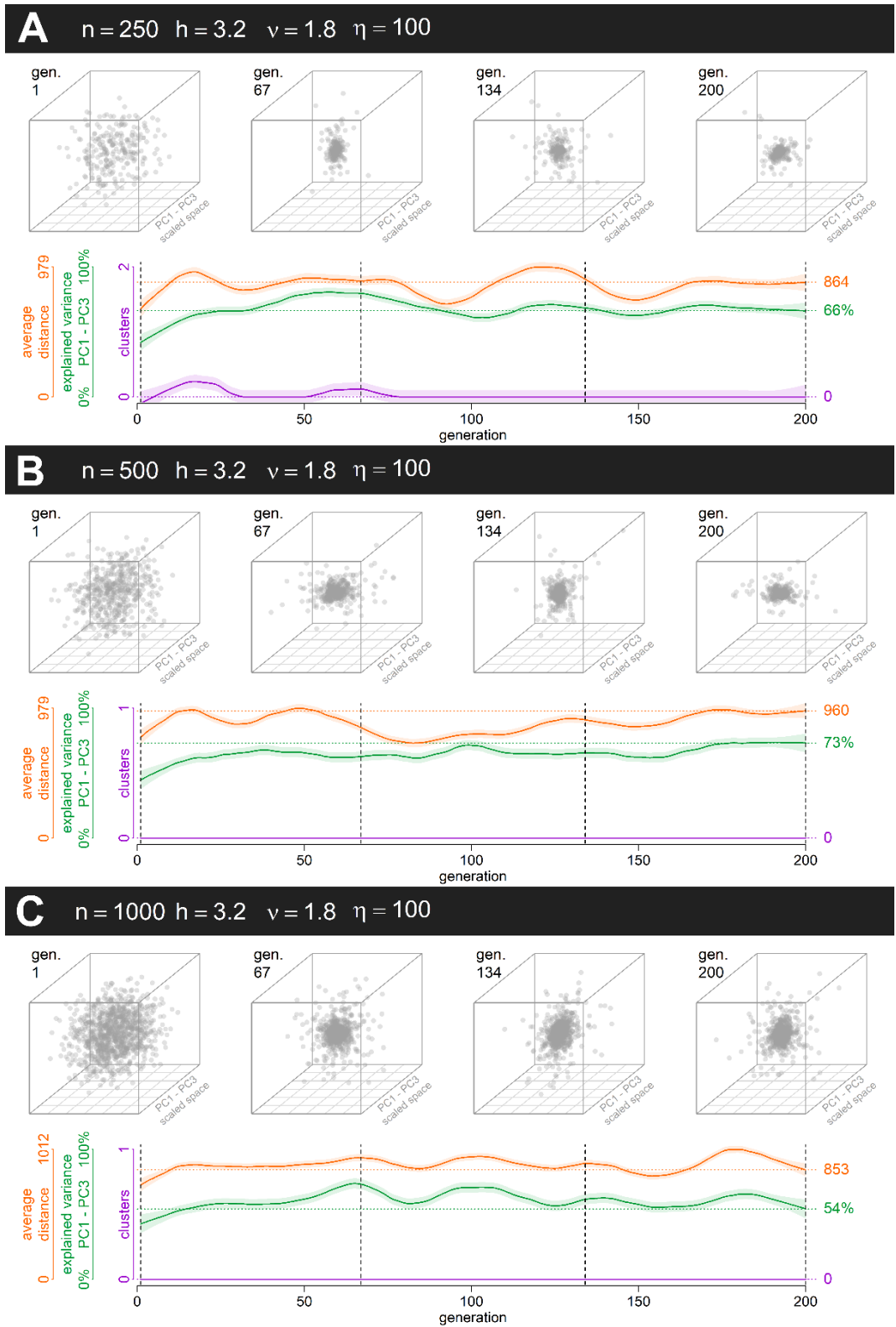

**Figure S7. Example simulations with a single distinct cluster due to a dominant influence of Galton-Pearson inheritance ($\eta = 100$) despite high homophily ($h = 3.2$) and relative offspring variance ($\nu = 1.8$); one simulation per small (A: n = 250), medium (B: n = 500), and large (C: n=1,000) population.** (See supplementary animations S7A-S7C).

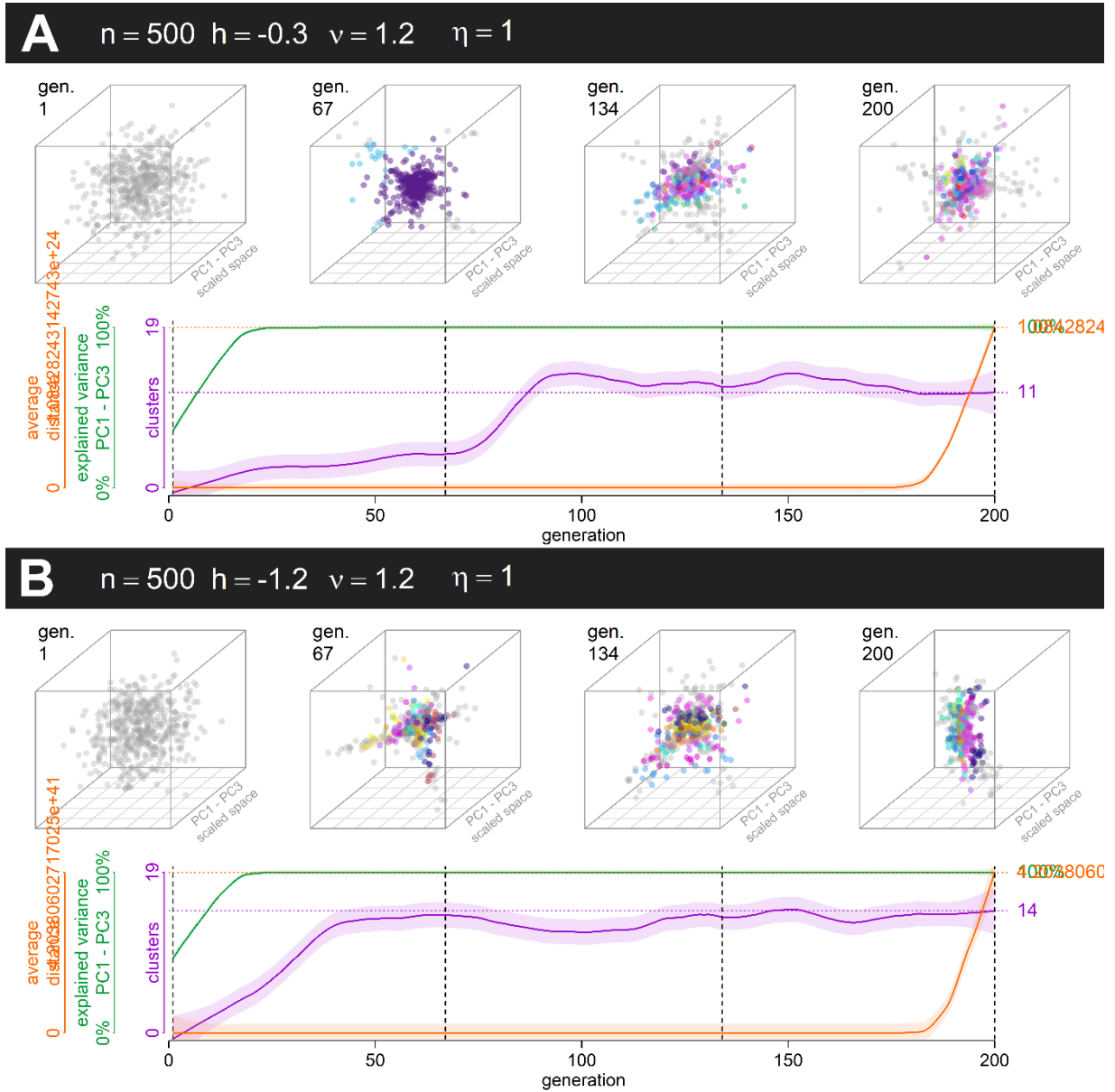

**Figure S8. Example simulations showing what happens when homophily is negative (A: h = -0.3, B: h = -1.2) and relative offspring variance is higher than 1 (v = 1.2).** Due to the absence of stabilising selection, we see a rapid spread along a single effective dimension of the culture space (See supplementary animations S8A and S8B).

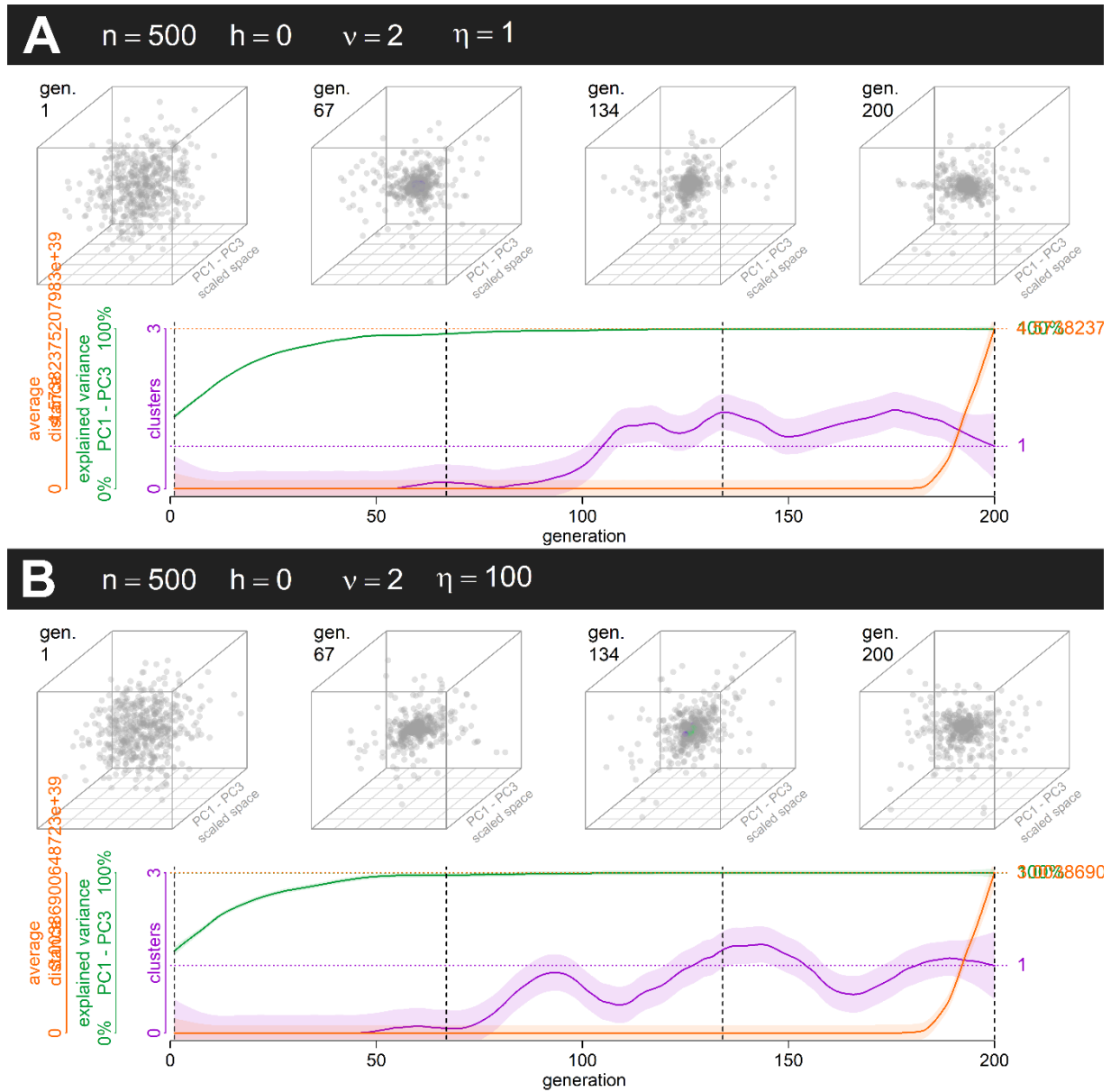

**Figure S9. Example simulations that show a rapid spread along a single effective culture space dimension despite random assortment.** This explosion is driven by a high relative offspring variance ($v = 2$). In such case, the role of constant offspring variance (A: $\eta = 1$, B: $\eta = 100$) is negligible. (See supplementary animation S9A and S9B).

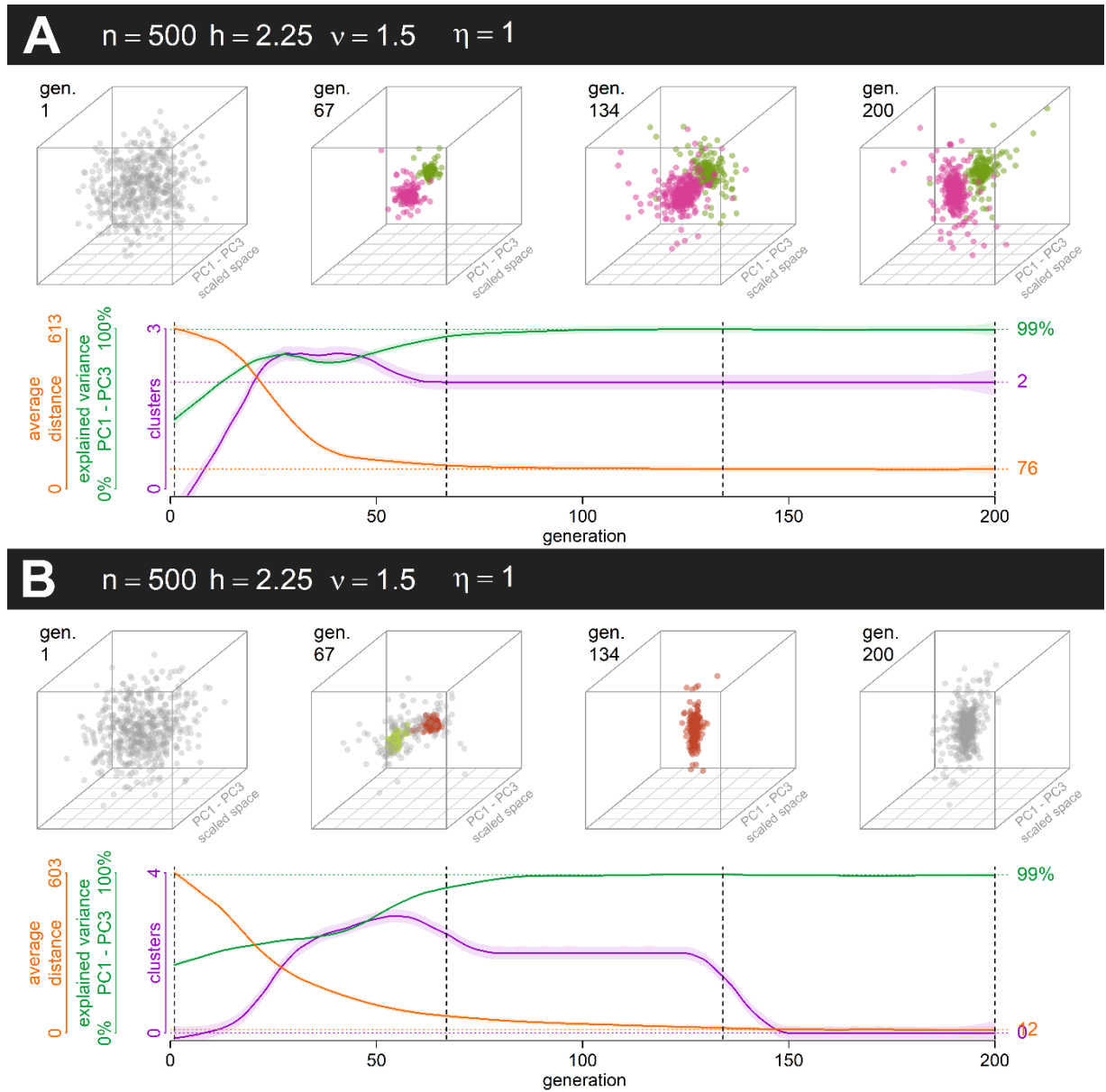

**Figure S10. Example simulation of a system with parameter combination that leads to a general loss of cultural variance but has a chance to form a small number of polarised clusters.** Typically, such parameter combinations lie on a 'black bridge' apparent on panels for $\eta = 1$ and $\eta = 10$ in FigureS1. The dimension reduction is due to the emergence of just two stable clusters. This 'polarisation of the population' is typically reflected in a PC1 that monopolises most of the variance. In A, two distinct clusters emerged and helped preserve some of the overall variance (see also the high proportion of variance explained by PC1–PC3). That, however, was not a universal outcome. All variance was lost where the population failed to form stable clusters (regardless of identical initial conditions). In such a case, the population collapsed into a single point (See supplementary animations S10A and S10B).

# Supplement S3. Results of a model where agent distance is normalised after each step

The model presented in the main manuscript has a vast advantage of being simple and straightforward. It does not introduce any additional assumptions besides the pairing and inheritance algorithm. The simplicity does, however, have some side effects. For example, for many combinations of parameter values (when the configuration of agents' position in the trait space is unconstrained), the basic system leads to one of the two extreme configurations: (1) All positions collapse into a tiny area within the trait space, where the pairing is as good as random (in case both $\eta$ and $\nu$ are small), or (2) the configuration explodes to infinity along a single effective dimension of the trait space (in case the $\nu$ is too large or $\nu$ is positive and $h$ negative). Both extreme outcomes can be avoided if the entire configuration is normalised without distortion to maintain a constant average distance between agents in the trait space. This allows us to focus on changes in the relative configuration of agents without the interference of variability loss or explosion. (Yes, a loss of all cultural variability may be a real possibility for the natural phenomena in question, but we can leave it aside for the moment being.) Most findings, however, remain, unchanged regardless of the model.

In the results below, the methods are the same as in the main manuscript, only after each generation, the point coordinates are normalised to the average distance of 436, which is approximately the average distance between normally distributed points in 10 dimensions with sd=100 along each dimension.

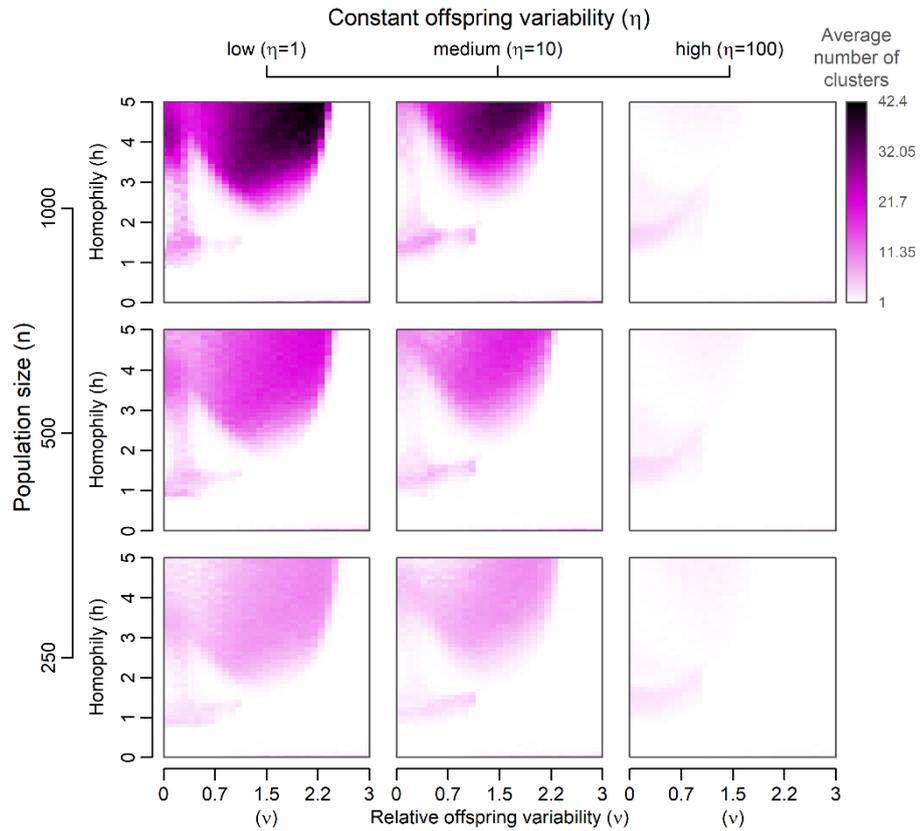

**Figure S11. A graphical summary of the tendency to form subcultures after 200 model generations in a system with normalized average distance**. The points in the 10-dimensional culture space were normally distributed across all dimensions at the beginning of each simulation run, and 100 simulation runs were executed for each parameter combination.

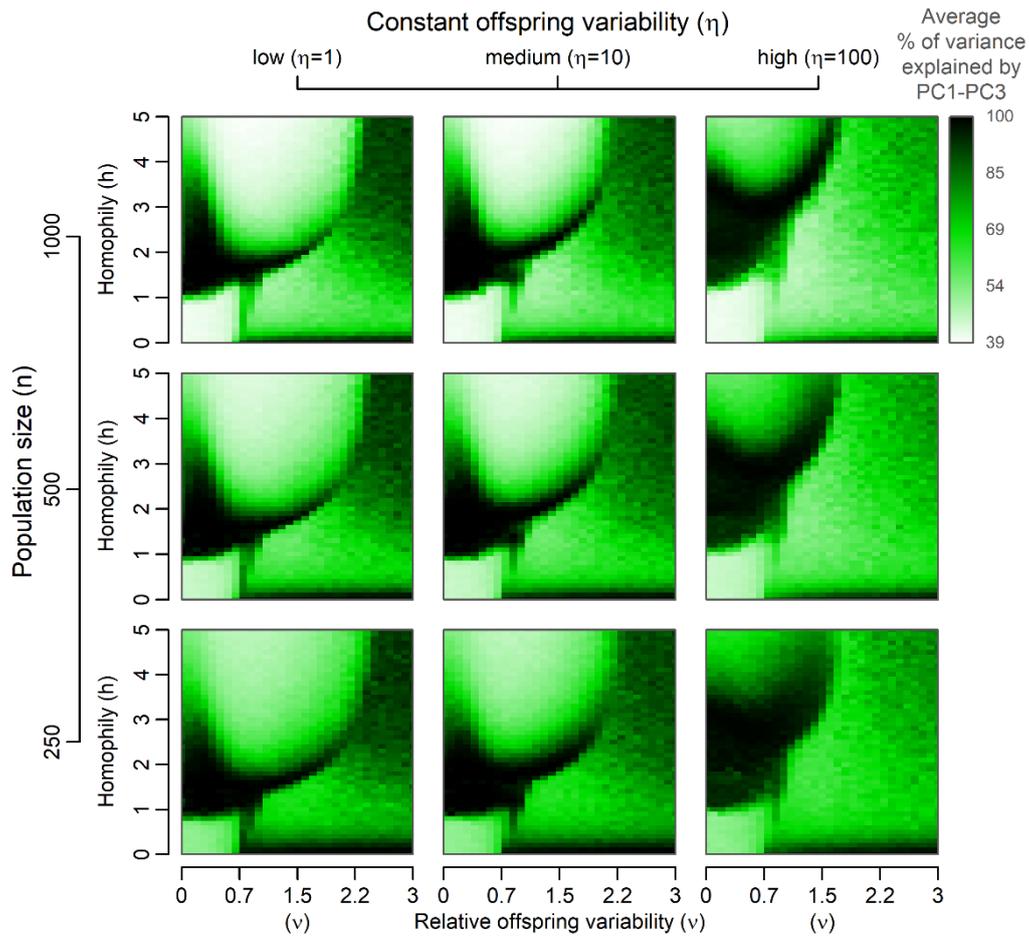

**Figure S12. A graphical summary of expected reduction in the number of effective dimensions after 200 model generations in a system with normalized average distance.** The same set of simulation runs as in Figure S11 was used to generate the image. PC1–PC3 stand for the first three Principal Components. See the explanation of the dimension reduction in a system with large $\eta$ and small $\nu$ in the caption of Figure S14.

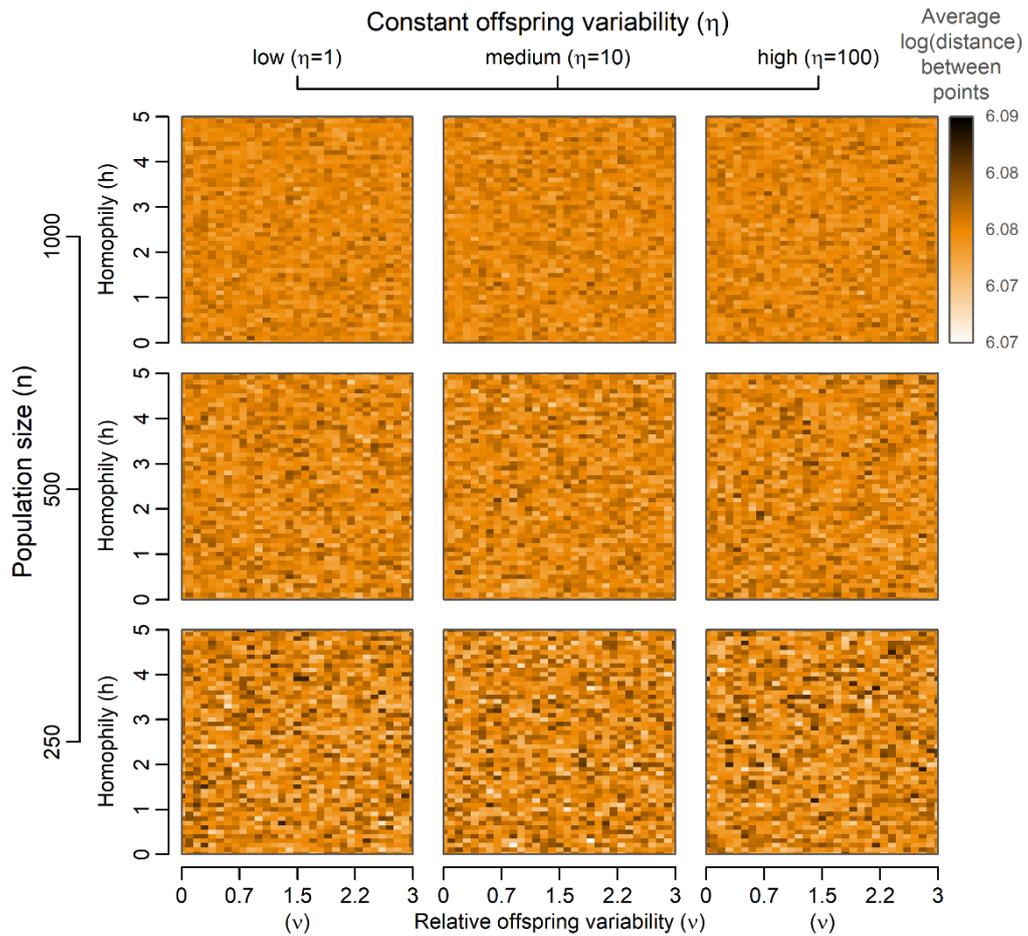

**Figure S13. A graphical summary of the expected average distance between agent positions in a culture-space after 200 model generations in a system with normalized average distance.** The same set of simulation runs as in Figure S11 was used to generate the image.

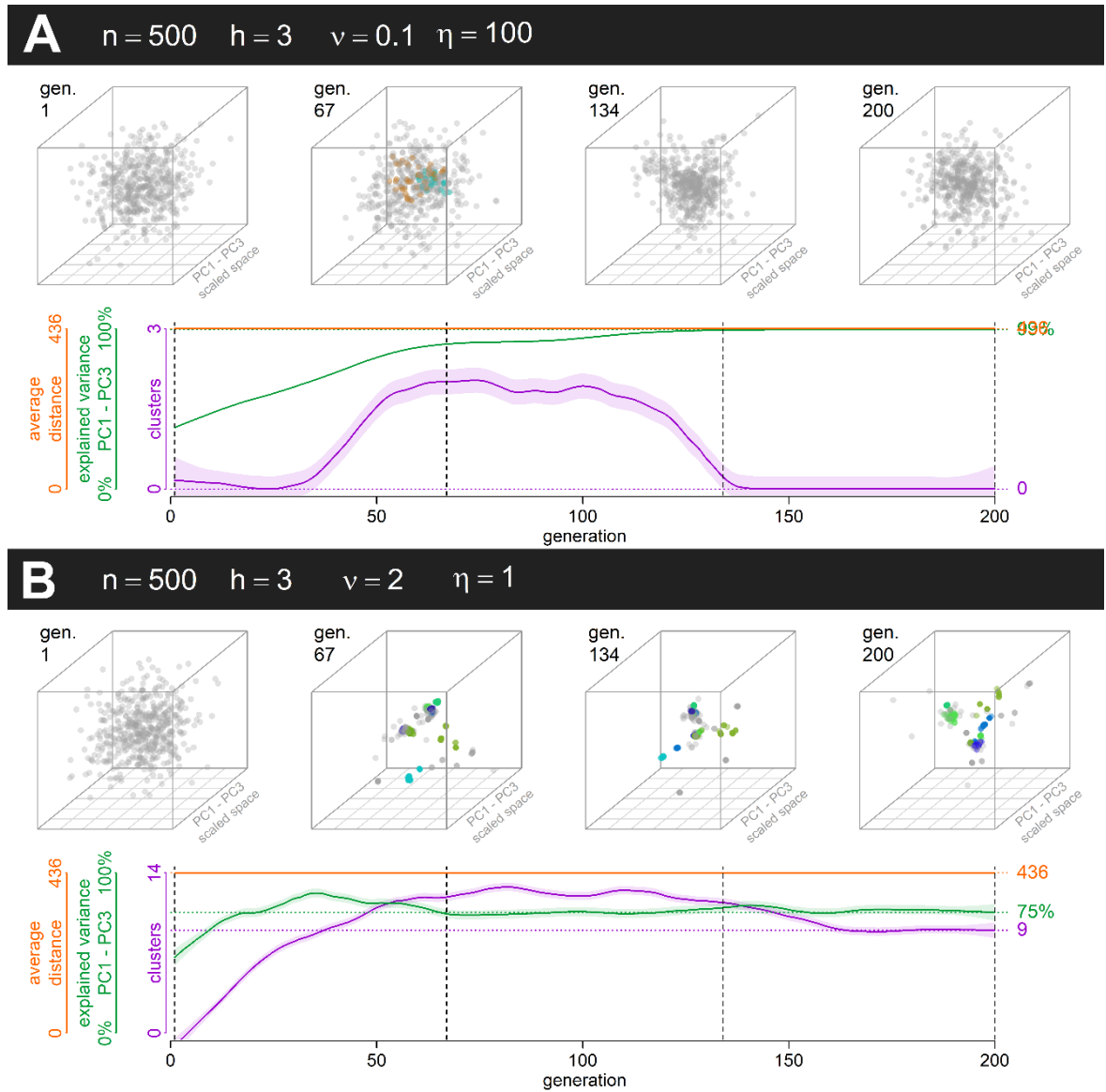

**Figure S14. Two simulation runs in a system with normalized average distance, one with strong influence of Galton-Pearson inheritance (A) and one with strong influence of PVDI (B).**

For detailed description see the caption of Figure 3 in the main article. Simulation run parameters were chosen to match those of Figure 3 Equivalents of Figures S3-S10 are not included for brevity, but can be easily obtained using the code at https://osf.io/pvyhe/?view_only=a79f0d07847f45cab069a8b8f09f258b. Notice the clustering tendencies and dimension reduction in the system with large influence of GP inheritance and homophily. The polarization is due to inflation of the originally small differences. If the homophily is large and there is normalization after each simulated generation, distances change proportionally to their magnitude. Similarly as in a system with PVDI and without normalization (see Figure 3B), larger distances get larger, which may cause polarization.

# Supplement S4. Results with a random variation along all dimensions of the trait space

One could argue that a model where all random variation is realised along the vector linking parental positions is not realistic. It is trivial to modify the inheritance part of the model summarised in equation 1.3 of the main manuscript and add random variation along the D dimensions of the trait space (S4.1). The assumption of a larger variation (proportional to the parental distance for $v \neq 0$) along the vector connecting parental positions is preserved.

$$A_o = \mu(A_{p1}, A_{p2}) + \frac{A_{p1} - A_{p2}}{\|A_{p1} - A_{p2}\|} N\left(0, \eta^2 + v^2 \frac{\|A_{p1} - A_{p2}\|^2}{4}\right) + X \qquad (S4.1)$$

The results after this slight modification: $X = (N(0,\xi)_1, N(0,\xi)_2, \ldots, N(0,\xi)_D) \in \mathbb{R}^D$, $\xi$ being a random phenotypic mutation of the offspring independent of parental positions, for $\xi = 1$, are included below.

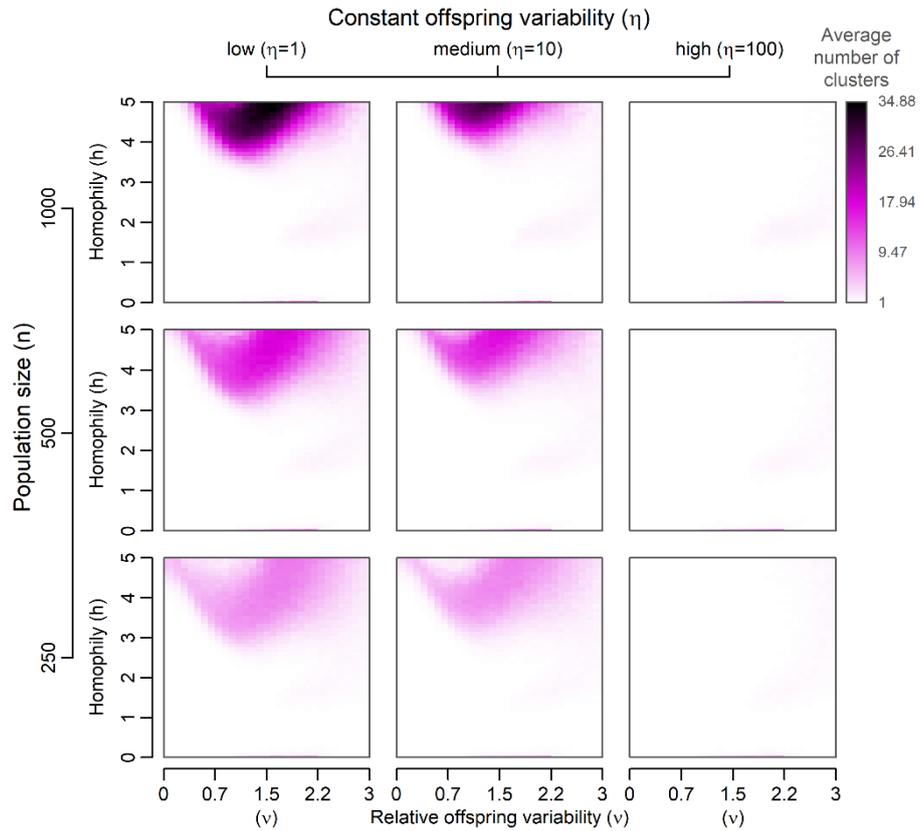

**Figure S15. A graphical summary of the tendency to form subcultures after 200 model generations in a system with an additional normal noise**. The points in the 10-dimensional culture space were normally distributed across all dimensions at the beginning of each simulation run, and 100 simulation runs were executed for each parameter combination.

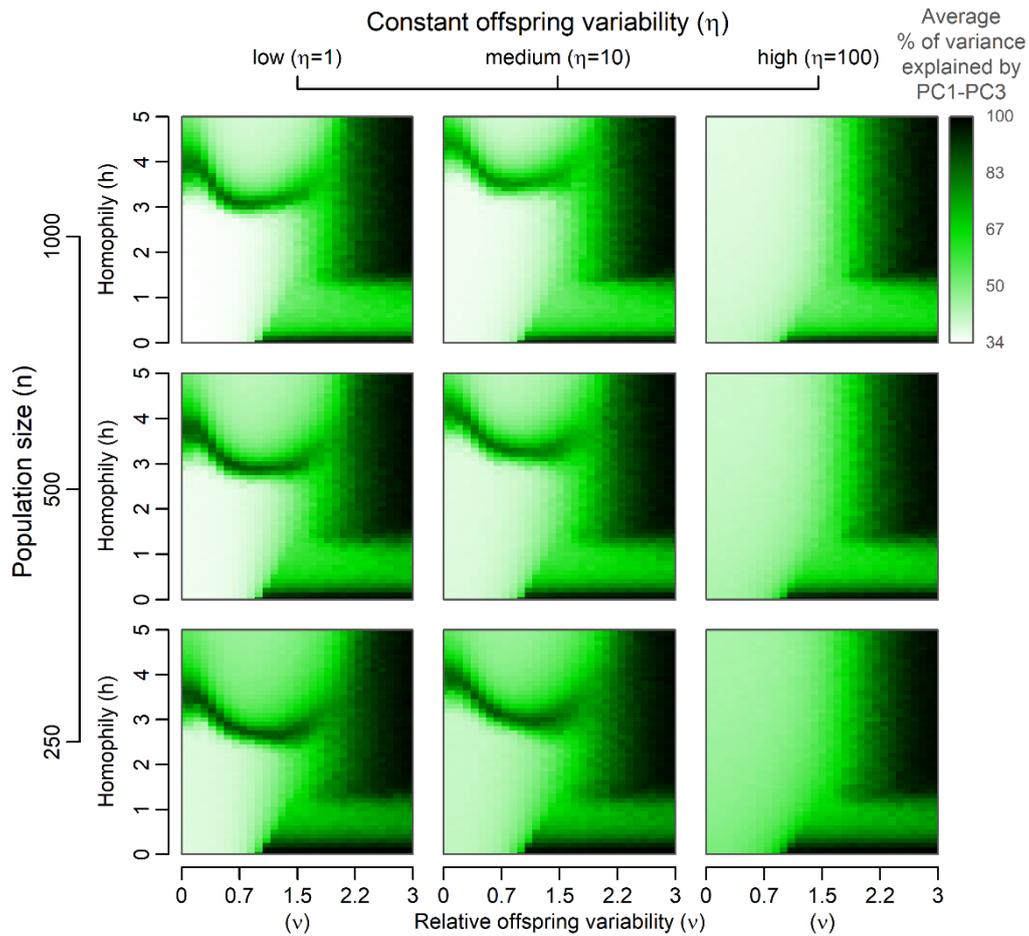

Figure S16. **A graphical summary of expected reduction in the number of effective dimensions after 200 model generations in a system an additional normal noise** The same set of simulation runs as in Figure S15 was used to generate the image. PC1–PC3 stand for the first three Principal Components.

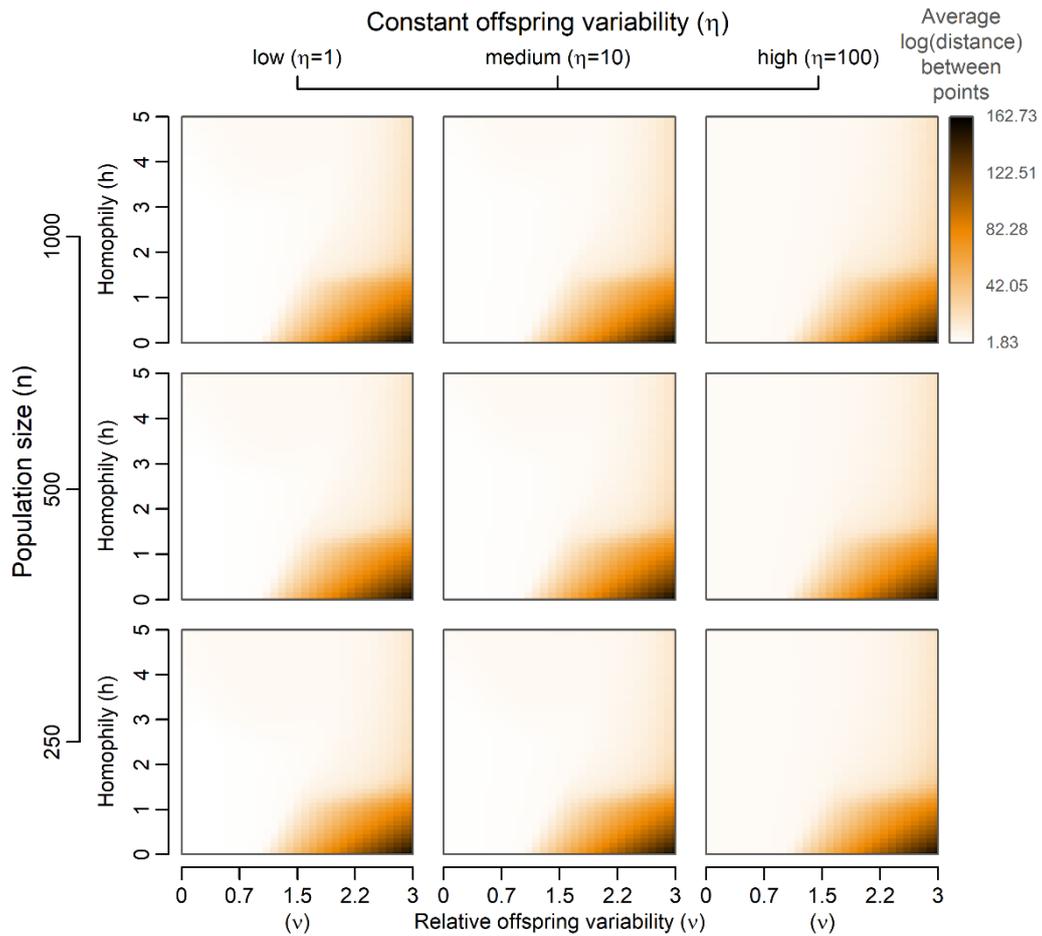

**Figure S17. A graphical summary of the expected average distance between agent positions in a culture-space after 200 model generations in a system with additional random noise.** The same set of simulation runs as in Figure S15 was used to generate the image.

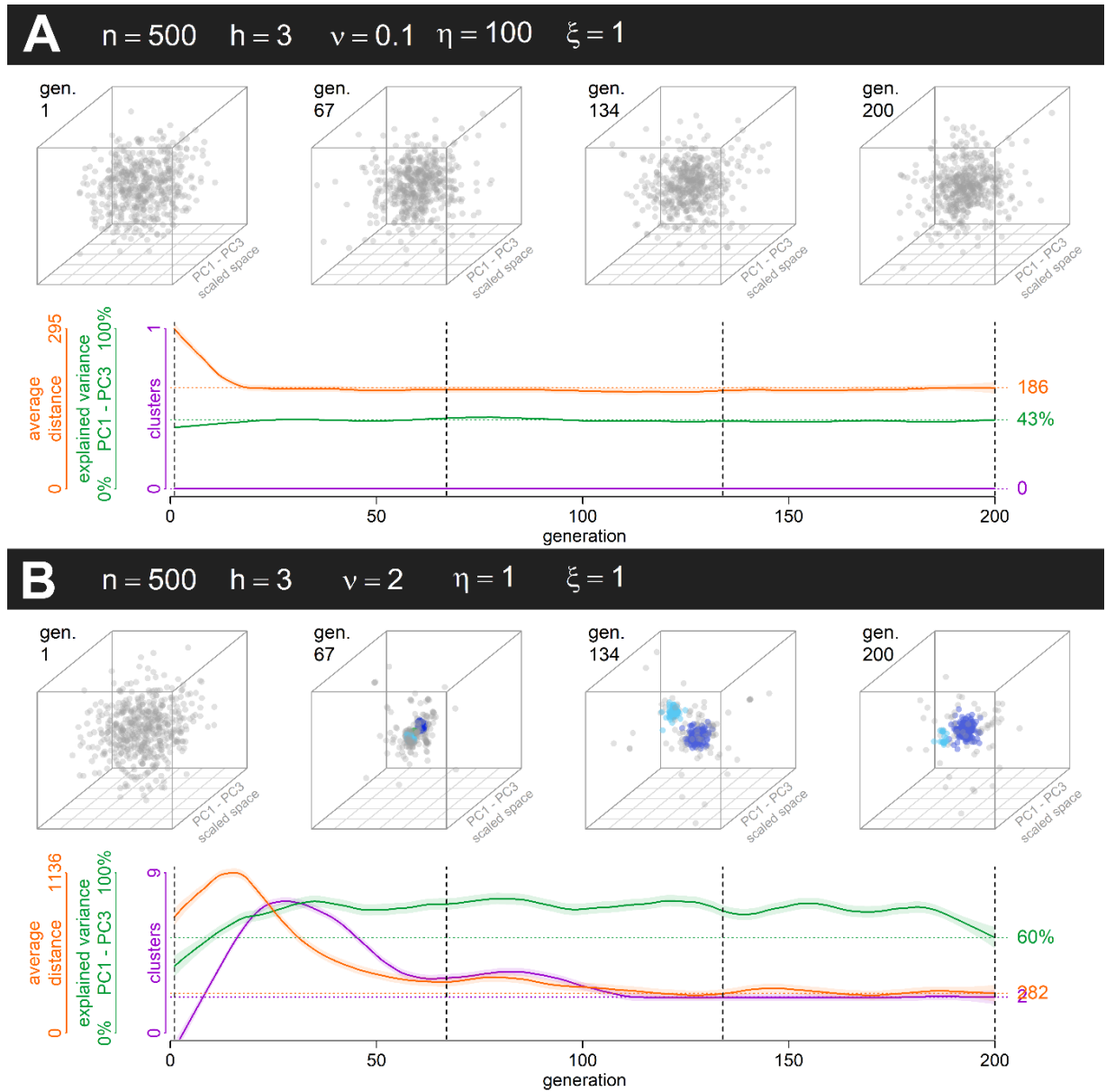

**Figure S18. Two simulation runs in a system with an additional random noise, one with strong influence of Galton-Pearson inheritance (A) and one with strong influence of PVDI (B).**

For detailed description see the caption of Figure 3 in the main article. Simulation run parameters were chosen to match those of Figure 3 Equivalents of Figures S3-S10 are not included for brevity, but can be easily obtained using the code at https://osf.io/pvyhe/?view_only=a79f0d07847f45cab069a8b8f09f258b.